%% file: main.tex
\documentclass[conference]{IEEEtran}
\IEEEoverridecommandlockouts
\usepackage{cite}
\usepackage{amsmath,amssymb,amsfonts}
\usepackage{algorithmic}
\usepackage{graphicx}
\usepackage{textcomp}
\usepackage{url}
\usepackage{makecell}
\usepackage{xcolor}
\usepackage{eurosym}  
\def\BibTeX{{\rm B\kern-.05em{\sc i\kern-.025em b}\kern-.08em
    T\kern-.1667em\lower.7ex\hbox{E}\kern-.125emX}}

\usepackage{fancyhdr}
\begin{document}

\title{A Practical AI-Driven Strategy for Cell On/Off Switching under Adaptable QoS Constraints \\
\thanks{This work is supported by the Grant TRAINER-6G (PID2023-146748OB-I00) funded by MCIN/AEI/10.13039/501100011033 and by ERDF/EU, by the Smart Networks and Services Joint Undertaking (SNS JU) under the European Union’s Horizon Europe research and by the Open Call ECO-RAN under the EU-funded project 6G-SANDBOX (Grant Agreement No. 101096328). It also received support from the COALESCE-6G project (PID2024-163028OB-I00), funded by MICIU/AEI/10.13039/501100011033/FEDER, EU. The authors further acknowledge the support of the CERCA Programme of the Generalitat de Catalunya.}
}

\author{
    \IEEEauthorblockN{
        David Reiss\IEEEauthorrefmark{1}, 
        Miguel Catalan-Cid\IEEEauthorrefmark{2}, 
        Daniel Camps-Mur\IEEEauthorrefmark{2}, 
        Oriol Sallent\IEEEauthorrefmark{1}
    }
    \IEEEauthorblockA{\IEEEauthorrefmark{1}UPC, Spain. \{david.reiss, jose.oriol.sallent\}@upc.edu}
    \IEEEauthorblockA{\IEEEauthorrefmark{2}i2CAT Foundation, Spain. \{david.reiss, miguel.catalan, daniel.camps\}@i2cat.net}
}

\maketitle
\thispagestyle{fancy} 

\begin{abstract}
The rapid expansion of 5G networks has intensified concerns over their sustainability, as denser Radio Access Network (RAN) deployments have increased overall power consumption. Although numerous studies have examined energy-efficient cell on/off switching, few have focused on approaches capable of dynamically adapting to operator-defined Quality of Service (QoS) requirements. In this paper, we propose a Long Short-Term Memory (LSTM)–based strategy, trained using a dataset from a European Mobile Network Operator (MNO), that enforces both target throughput levels and outage-tolerance constraints. Unlike previous approaches, our model adapts to different QoS requirements by tuning a decision threshold at inference time, enabling operators to balance energy savings and service guarantees without retraining. Across an unseen week, the method attains 63–96\% of an oracle’s energy savings while largely meeting operator-specified constraints. We also provide CO\textsubscript{2} and OPEX estimates under representative scenarios to quantify potential operator benefits. 
\end{abstract}

\begin{IEEEkeywords}
Energy Efficiency, QoS, AI-Driven optimization, LSTM, 5G-RAN
\end{IEEEkeywords}

\input{introduction}
\input{solution_design}
\input{evaluation}
\input{conclusions}

\bibliographystyle{IEEEtran}
\bibliography{references}

\end{document}

%% file: introduction.tex
\section{Introduction}\label{sec:intro}

{Reducing carbon footprint of mobile networks is key to align with global sustainability goals. According to the 2025 GSMA white paper \cite{gsma}, transitioning towards greener operation has become a defining priority for the mobile communications industry. The report notes that 45 Mobile Network Operators (MNOs) worldwide have pledged to achieve net-zero emissions by 2050, with nearly half of them targeting 2040. In response, the research community has increasingly focused on the development of energy-saving mechanisms for 5G and beyond networks \cite{3GPP_esavings}. However, a central challenge is to ensure that these mechanisms preserve service quality, which is a fundamental requirement for operators. This motivates the design of strategies which explicitly manage the energy–QoS trade-off, such as rule-based heuristics and static strategies that leverage sleep modes and predefined operational states (e.g., power amplifier muting, carrier shutdown, Multiple Input Multiple Output (MIMO)-chain deactivation) to curb Radio Access Network (RAN) power consumption \cite{esaving_techonologies}.}

{Advanced AI/ML-based mechanisms are key to applying dynamic on/off strategies. The O-RAN architecture eases the integration and deployment of AI/ML workflows, and therefore is being extensively used to develop ML-driven solutions \cite{BeGREEN_D42}. Indeed, the O-RAN Alliance has released different technical specifications covering cell on/off switching use cases guided by traffic predictive tasks executed at both real-time and non-real-time domains \cite{oran-esaving-latest}. In \cite{greener-ran-operation}, an ML-driven traffic forecasting approach is used to enhance energy savings through intelligent control of BS sleep modes. Several well-known algorithms are compared: Block Linear Regressor, Artificial Neural Networks (ANNs), and Long Short-Term Memory (LSTM). Results show that all approaches achieve similar performance, attaining 10\% to 40\% of energy savings. }

{In addition, effective QoS awareness is essential, as maintaining user experience is a fundamental network requirement. In \cite{advanced-sleep-modes}, authors present an energy-efficient radio unit (RU) control via Advanced Sleep Modes. The RU is shown to contribute up to 90\% of the total energy consumption, noting that  Power Amplifier Module (PAM) efficiency improves with Physical Resource Block (PRB) utilization. Accordingly, the authors present a lightweight ML-driven framework, used to control and coordinate radio scheduling through an xApp, which adapts in near real-time to the network conditions. Moreover, the authors analyze the impact of imposing specific QoS constraints, which are based on UE-experienced delays caused by the PRB re-scheduling algorithm. Results show how energy reductions range from 15\% to 72\% according to the QoS requirements, highlighting the impact of imposing severe constraints. }

{In \cite{drl-lstfm-federated}, a Deep Reinforcement Learning (DRL) agent decides BS switching by predicting UE locations through a Federated LSTM (F-LSTM) system. Forecasting UEs trajectories allows the DRL agent to estimate the impact on QoS degradation caused by handover operations to sub-optimal BSs. The simulated results show that energy efficiency is increased by 36\% with respect to a full activation scenario while minimizing QoS degradation. A mobility dataset was used to train models further used to predict UE trajectories. However, no real-world network data was used, which is essential for realistic studies and for training and validating models under conditions simulations may not capture.}

{Moreover, in \cite{gnn-switching} a Graph Neural Network (GNN) based cell switching approach is tested over an ultra-dense heterogeneous network. By using an open dataset from Telecom Italia, authors aim at achieving energy savings while preserving QoS, measured as user throughput. Results show that the proposed approach achieves 10.4\% of energy efficiency gains while preserving 99.6\% of users QoS. Moreover, the performance achieves 75.7\% of the optimal solution, obtained by running an exhaustive search algorithm. }

{Although the above literature provides strong and consistent results, most studies characterize the energy savings with respect to minimum or sufficient QoS requirements, but do not provide tools that MNOs can use to configure the energy saving strategies according to explicit QoS constraints. According to the study we performed in \cite{previous-paper}, where we used a real dataset from an European MNO to quantify the energy-QoS trade-off, potential energy savings exist even under stringent QoS constraints. By utilizing an oracle-strategy (also used in this work to benchmark the results), we searched for periods of time during which the 5G traffic could be balanced to the 4G cells found in the same site and sector. The oracle strategy also provided us an estimation on the throughput provided during the switch-off periods, which is key to develop a QoS-aware energy saving strategy. In this paper, we extend the work we presented in \cite{mswim-paper}, where we used an XGBoost Classifier to develop a 5G cell on/off switching strategy that guaranteed 15 Mbps during the switch-off periods, and was able to bound the outage decisions. Although the proposed methodology reported good performance, it required either to be re-trained each time the throughput requirement was modified, or to have several models ready to be used for each QoS requirement. This presented a substantial disadvantage in terms of scalability and model maintenance.}

{The main contribution of this work is the development of a Neural Network (NN)-based cell on/off switching strategy capable of seamlessly adapting to any changes in the defined QoS constraints. Building on prior evidence that cell switch-off can yield sizable savings under QoS constraints, we now target operator-controllable enforcement at inference time. Unlike our earlier classifier-based study, where policy compliance required class-ratio tuning per model \cite{mswim-paper}, in this paper we adopt a temporal LSTM and post-hoc thresholding to meet different constraints without retraining, which is more suitable for live operations. In addition, the proposed strategy enables MNOs to flexibly adapt QoS requirements to ensure that end users will not be affected by the energy saving mechanisms. Moreover, we provide a high-level estimation on the potential operator's carbon footprint reduction and economic benefits for the operator resulting from the deactivation of underutilized 5G cells. }

{The paper is structured as follows. The next section presents the general methodology, the NN configuration, and the outage enforcement mechanism, all contributing to describe the proposed solution design. Section \ref{sec:evaluation} presents the performance evaluation across an unseen week of data, and a wide range of QoS constraints. Then, the results are mapped into carbon footprint and economic expenses reduction, showing the potential benefits from an operator's perspective. Finally, in Section \ref{sec:conclusions} we present the conclusions and the future work. }

%% file: solution_design.tex
\section{Solution design}\label{sec:solution}

{In this section we present the design of the QoS-aware energy saving strategy. Subsection A details the general methodology and the generation of the training, testing, and evaluation datasets. Subsection B presents the neural network (NN) configuration, and details the mechanism to make our strategy aware of the two defined QoS dimensions. Then, subsection C uses the testing week to characterize the effect of decision threshold biasing on the outage decisions and missed opportunities. }

\subsection{General methodology}

{We reuse the real-network dataset described in our prior studies. Full dataset details are in \cite{previous-paper}\cite{mswim-paper}; here we summarize only elements needed for reproduction. We have access to a full month of real network data provided by a European MNO. The dataset has a granularity of 15 minutes and provides a long list of 4G and 5G KPIs. We target a concrete area of an urban scenario which has a total of 24 deployed sites. As usual, each site is divided in three sectors, on which 4G and 5G cells from different carriers are collocated, leading to a total of 70 5G cells and 3 to 5 active 4G cells per sector (depending on the site).} 

{In accordance with our previous work, the designed energy saving strategy needs to enhance control along two distinct QoS dimensions: (i) target throughput level, which defines the minimum throughput $T$ to be guaranteed during the 5G cells' switch-off periods in the 4G cells, and (ii) outage tolerance constraint, which bounds the maximum outage decisions (incorrect switch-off choices that lead the end user to experience service degradation) to a defined threshold $\gamma$. Concretely, we define six throughput levels: 0 Mbps (i.e., minimum to guarantee connectivity), 5 Mbps, 10 Mbps, 15 Mbps, 20 Mbps, 25 Mbps, and three outage constraints: 10\%, 5\%, 3\%.}

{The first step is to generate a dataset which contains information on the switch-off opportunities across all defined throughput levels over the full month of data. This is done by running the oracle strategy, which analyzes the PRBs of the cells in each sector and site, balances the 5G load across 4G cells to simulate a switch-off phase, and, using a regressor that leverages the correlation between 4G cell utilization level and average throughput per UE, decides whether the 5G cell can remain off while still meeting the required throughput level. We refer to \cite{previous-paper} for more details on the oracle methodology and performance. This methodology produces the dataset shown in Table \ref{tab:dataset}. }

\begin{table}[h]
\centering
\caption{Training Dataset}
\begin{tabular}{|c|c|c|c|c|c|c|}
\hline
\makecell{Time} & \makecell{5G\\ cell\\ load \\(\%)} & \makecell{4G \\cell 1 \\load \\(\%)} & {···} & \makecell{4G \\cell 5 \\load \\(\%)} & \makecell{Throughput\\level, $T$ \\ (Mbps)} & \makecell{Decision}\\
\hline
$t_1$ & 54.3 & 45.2 & {···} & 67.8 & 10 & 1  \\
\hline
$t_2$ & 37.8 & 40.1 & {···} & 50.1 & 10 & 1  \\
\hline
$t_3$ & 24.2 & 35.1 & {···} & 42.5 & 10 & 0  \\
\hline
\end{tabular}
\label{tab:dataset}
\end{table}

{The dataset is generated for all six defined throughput levels (not just the one shown in the table). The load of the cells and the targeted throughput level are the inputs to the model, and the "Decision" column is the labeled output, which indicates with a "0" if the 5G cell can be switched off while maintaining the input throughput requested, and with a "1" if it must be kept on. The first two weeks of the generated dataset are used to train the NN, the third week is used to test the effectiveness of the defined outage enforcement mechanism (next section), and finally, the last week is used for the performance evaluation of the proposed strategy. }

\subsection{NN configuration and outage enforcement mechanism}

{By design, LSTM NNs have been widely used to address different tasks concerning time-series and sequentially structured data, where modeling temporal dependencies is essential. The ability for LSTMs to handle sequence-to-sequence mappings has also been exploited in the energy savings field, where forecasting future network behavior becomes a key requirement \cite{lstm-prediction}. In our case, the 5G cell on/off switching decision must be taken using the utilization KPIs of the different 4G and 5G cells obtained from the MNO's dataset, which inherently exhibit time-series characteristics. These KPIs reveal daily periodic patterns, with high utilization during midday hours, and low utilization during nighttime. Since the switch-off opportunities are influenced by the input loads, using an LSTM enables the NN to effectively capture temporal dependencies, such as increasing/decreasing utilization trends (e.g., night-to-day transitions), or static utilization (e.g., during peak and nighttime hours). To build the neural network, we use the Keras library\footnote{https://keras.io/api/layers/}, which has been extensively used to build Deep Learning (DL) solutions. Table \ref{tab:lstm} shows the configuration parameters of the proposed NN.}

\begin{table}[h]
\centering
\caption{NN Configuration}
\begin{tabular}{l c}
\hline
\textbf{Parameter} & \textbf{Value} \\
\hline
Input features      & 7 \\
Sequence length      & 10 \\
LSTM layers          & 1 \\
LSTM units           & 32 \\
LSTM activation      & tanh \\
Dropout rate         & 0.1 \\
Dense output units   & 1 \\
Dense activation     & Sigmoid \\
Loss function        & Binary Cross-Entropy \\
Optimizer            & Adam \\
Batch size           & 16 \\
Epochs               & 20 \\
Validation split     & 0.1 \\
\hline
\label{tab:lstm}
\end{tabular}
\end{table}

{An implicit objective when defining an energy saving strategy is to ensure that the employed models exhibit low computational cost, therefore minimizing the energetic footprint associated with the training process. To this end, we configure our NN with only one 32-unit LSTM layer followed by a one-unit dense layer. Since the problem is formulated as a binary classification task, we use the binary cross-entropy as the loss function, and the Sigmoid as the output dense layer activation function. Thus, the output of the NN is on the $[0, 1]$ range and can be understood as $P(y=1)$, i.e., the probability of switching on. Moreover, to capture recent network behavior, we input the last 10 timesteps of the seven input features (equivalent to 2.5 hours of activity at 15-minute intervals), which is sufficient to represent increasing, decreasing, or static trends. Based on these inputs, the NN is tasked with predicting the future on/off decision one timestep ahead. In addition, training is performed with a batch size of 16, meaning that the LSTM parallelizes 16 samples (each consisting of 7 features $\times$ 10 time steps) before updating weights. This batch size corresponds to a 1/6 fraction of a complete period of the utilization KPIs, which have a daily-periodic pattern corresponding to 96 samples recorded every 15 minutes.}

{Up to this point, the NN has been made aware of the target throughput level $T$, which is included in the training feature set. However, no mechanism has yet been introduced to enforce outage constraints. This can be addressed in two different ways: }

\begin{itemize}
    \item {Custom loss function: instead of employing a standard binary cross-entropy loss function, a custom one is defined so that an extra penalty is introduced to make negative decisions ("0" or off) more costly, therefore bounding outage decisions. This biases the model during the training process. }
    \item {Decision threshold $\rho$: the NN produces an estimated probability $P(y=1)$, to which a decision threshold $\rho$ is applied such that the predicted label $\hat{y}$ is assigned to class "1" if $P(y=1) \geq \rho$, and to class "0" otherwise. Setting $\rho$ close to zero biases the model outputs towards positive decisions, therefore bounding outage decisions. }
\end{itemize}

{Although defining a custom loss function makes the model to be correctly biased, it requires re-training whenever a new bias level is desired (i.e., each time $T$ or $\gamma$ is modified). Moreover, if multiple bias configurations are needed, a separate NN must be trained for each case, which presents the same scalability issues as our previous approach \cite{mswim-paper}. By contrast, biasing the model outputs through $\rho$ allows training a single model once. Then, different QoS constraints can be guaranteed by simply adjusting $\rho$. }

{The NN exhibits high accuracy after training. Specifically, when using a default setting of $\rho = 0.5$, we get an average accuracy across throughput levels of 91.3\% (measured as the ratio of true positives and true negatives to the total number of decisions), with a maximum of 94.4\% at 0 Mbps, and a minimum of 88.3\% at 10 Mbps. Recall that different throughput levels have different switch-off opportunities, and therefore different class distributions (i.e., varying proportions of “1” and “0” labels in the generated dataset), suggesting that those results could be further improved by fine-tuning $\rho$ at each throughput level. }

\subsection{Effect of $\rho$ setting on NN performance over testing week}

\begin{figure}[t]
    \centering    \includegraphics[width=0.45\textwidth]{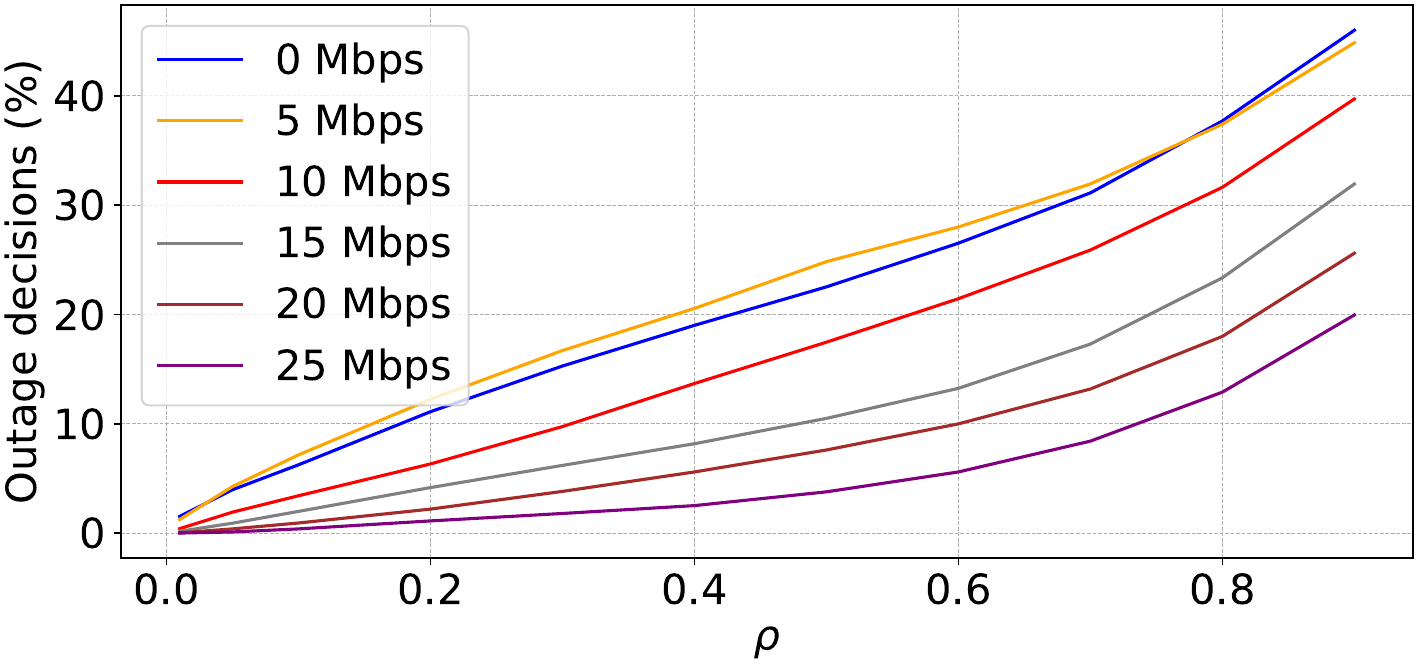}
    \includegraphics[width=0.45\textwidth]{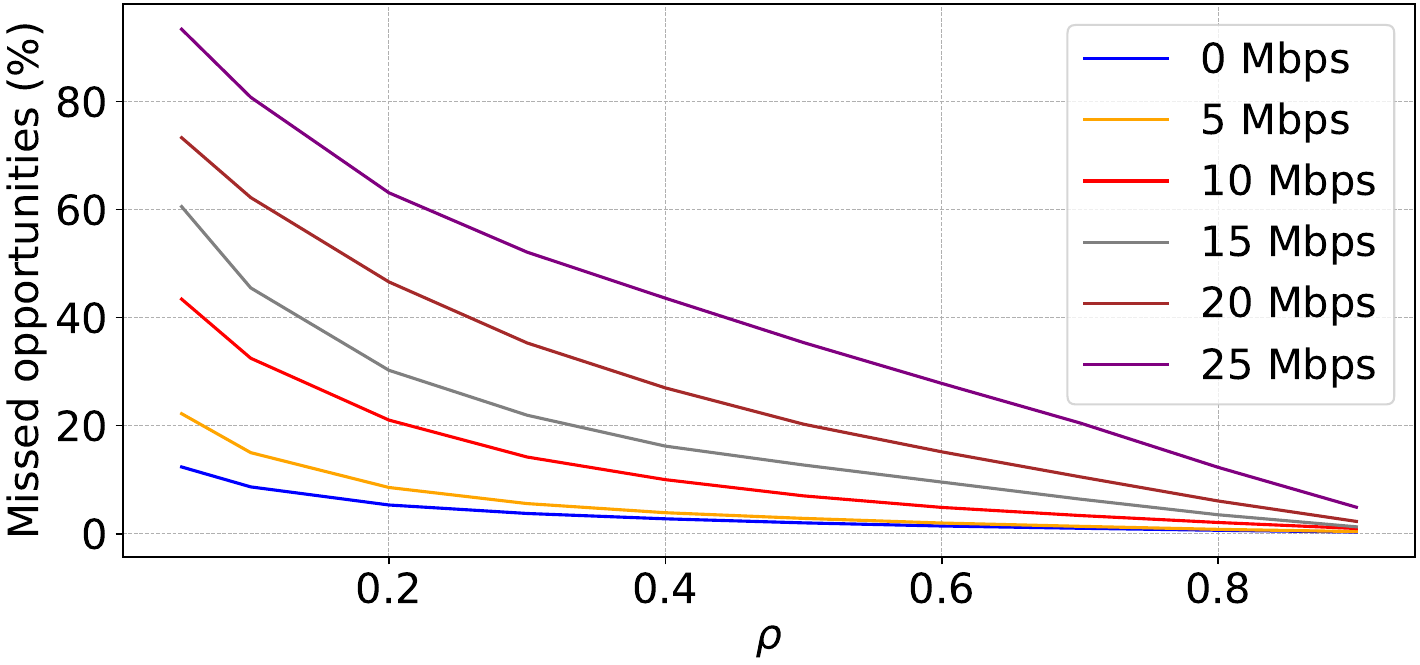}
    \caption{Missed opportunities and outage decisions trade-off.}
    \label{fig:outage-missed}
\end{figure}

{In this subsection, we analyze the effect varying $\rho$ has on the outage decisions and the missed opportunities across all throughput levels. The evaluation is performed using the third week of data, which was not included in the training process, and is unseen for the model. The results are presented in Figure \ref{fig:outage-missed}. Here, an outage decision is an incorrect 'off' prediction, normalized by the total number of instances where the cell should have remained 'on' (ground truth 'on'). A missed opportunity is an incorrect 'on' prediction, normalized by the total number of true switch-off opportunities (ground truth 'off').}

{As expected, the results indicate that outage decisions increase with higher values of $\rho$, while missed opportunities decrease, therefore highlighting a clear trade-off between both output metrics. This behavior is common across all throughput levels, and reveals that reducing outage decisions will inevitably come at the cost of cutting the achievable savings. However, because the number of switch-off opportunities varies with the target throughput level, higher $T$ values result in fewer outage decisions but a greater proportion of missed opportunities. This outcome demonstrates that the $\rho$ setting required to satisfy a given outage threshold $\gamma$ is not uniform across throughput levels. Consequently, if outage decisions were to be bounded through a custom loss function, separate NNs would need to be trained for each throughput level. On contrary, when using the decision threshold approach we can use the same base model across all $T$ levels.}

\begin{figure}[t]
    \centering    \includegraphics[width=0.49\textwidth]{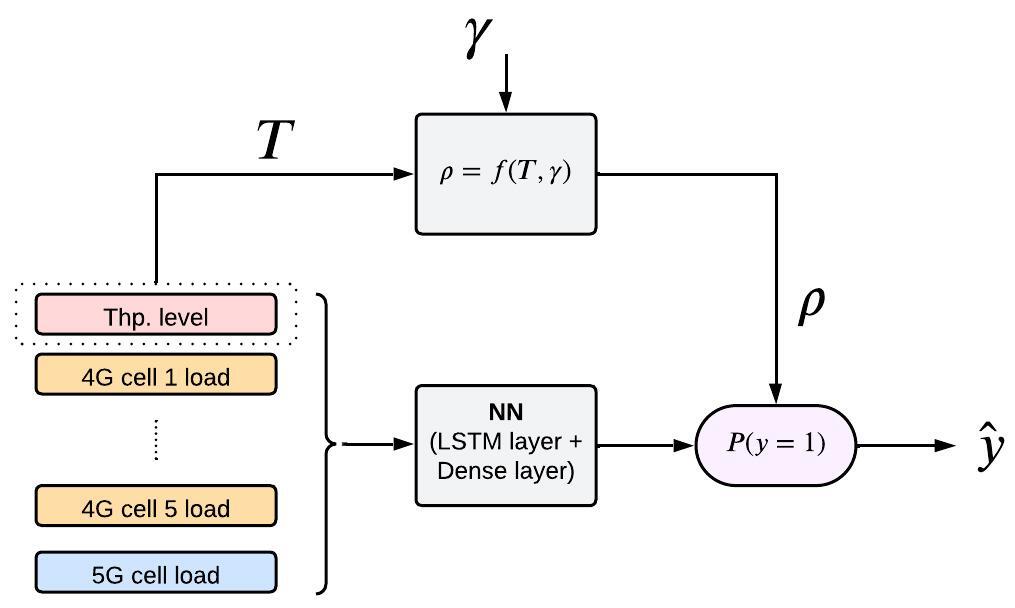}
    \caption{High level scheme of the energy saving strategy.}
    \label{fig:lstm}
\end{figure}

{Additionally, the results obtained over the testing week allow us to estimate $\rho$ for given $\gamma$ and $T$ by selecting the smallest $\rho$ value that results in an empirical outage of $\leq \gamma$. For instance, if we want to exploit energy savings while maintaining a throughput requirement of 10 Mbps, and keeping outage decisions below 5\%, we should set $\rho$ approximately equal to 0.2. This is the approach followed in the final design of the proposed energy saving strategy, illustrated in Figure \ref{fig:lstm}. Each time an inference is done, we forward the inputs through the NN while the QoS constraints are input to a function which, using a lookup table per $(T, \gamma)$, estimates the $\rho$ to be applied to the NN output probability. This way we are able to adjust the proposed solution to different QoS constraints in real-time, from one timestep to another if needed. }

%% file: evaluation.tex
\section{Performance evaluation}\label{sec:evaluation}

{In this section we evaluate the performance of the proposed strategy, where the NN and the $\rho$ determination is done at each inference according to the testing week results. Subsection A, presents the results obtained in terms of switch-off times and outage decisions. Then, subsection B maps those results into a CO\textsubscript{2} emissions reduction, and the associated economic benefits derived from the decrease in operational expenses. }

\subsection{Achievable savings and outage tolerance compliance}

\begin{figure}[t]
    \centering    \includegraphics[width=0.35\textwidth]{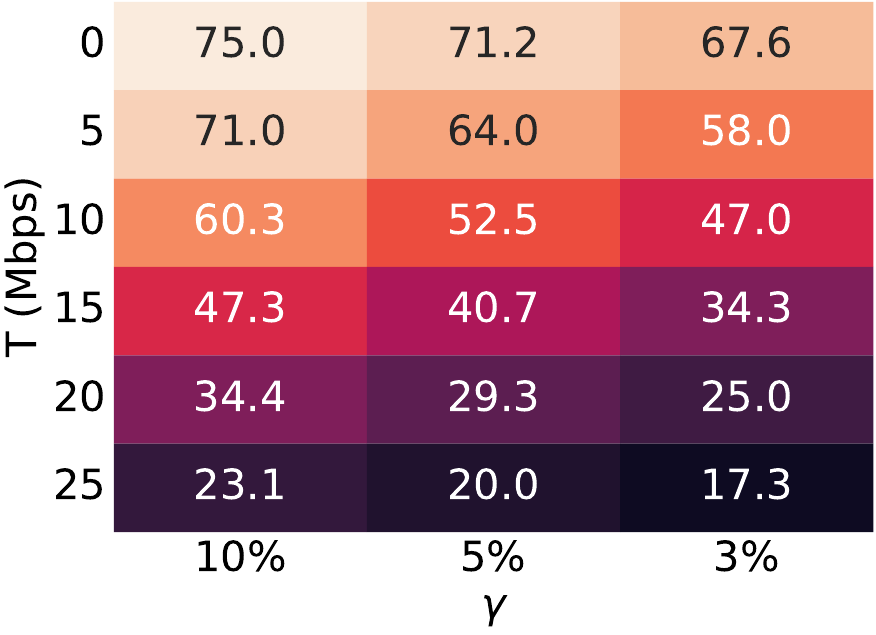}
    \caption{Switch-off times across analyzed cells over the evaluation week.}
    \label{fig:achievable-savings}
\end{figure}

{The methodology is straightforward: we evaluate the model for different tuples of $(T, \gamma)$, each of them requiring a different $\rho$, which is inferred by leveraging the results from the testing week. In order to minimize the impact on the missed opportunities, it is essential to select the minimum $\rho$ value that ensures the defined QoS constraints tuple. For example, when $\gamma$ is set equal to 3\%, we obtain $\rho$ equal to 0.03, 0.03, 0.09, 0.14, 0.25, 0.44 for each $T$ from 0 Mbps to 25 Mbps, respectively. The same process is repeated for all defined $\gamma$ thresholds. Figure \ref{fig:achievable-savings} presents the average switch-off time across the 70 analyzed cells  achieved by the proposed strategy during the evaluation week. }

{The trade-off is evident: the more stringent the QoS constraints in either of the two analyzed dimensions, the shorter the average switch-off times and, consequently, the lower the achievable energy savings. It is important to note that these results are obtained using a single model, demonstrating that the proposed strategy can seamlessly adapt to any imposed QoS constraint. Furthermore, when compared to the optimal solution, the approach exhibits very competitive performance. Specifically, the oracle strategy (which achieves savings without incurring any outage decisions) achieves energy savings of 78.6\%, 78.1\%, 67.8\%, 52.7\%, 39.3\%, and 27.3\% for throughput levels ranging from 0 Mbps to 25 Mbps, respectively. Thus, the proposed strategy always achieves between 63\% and 96\% of the oracle’s performance. }

\begin{figure}[t]
    \centering \includegraphics[width=0.44\textwidth]{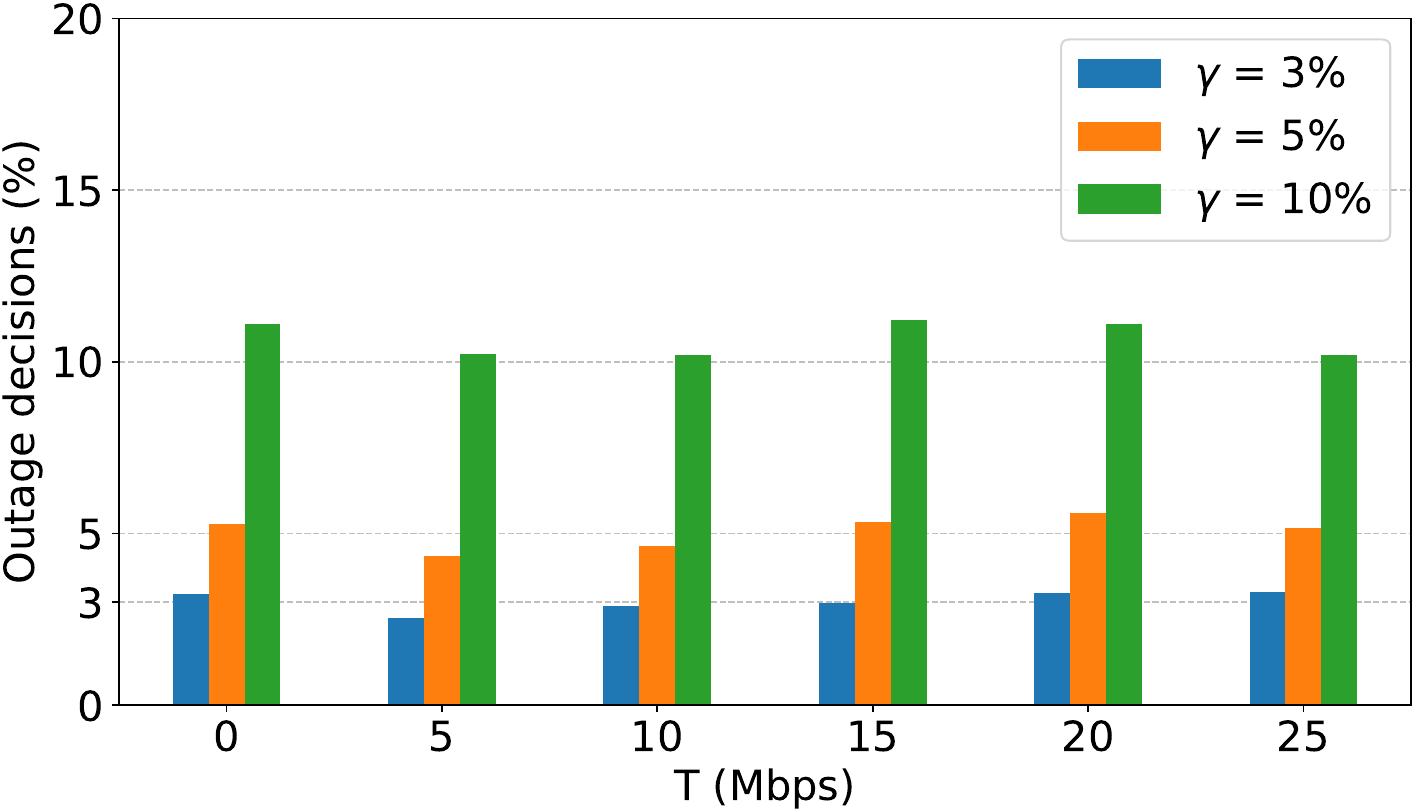}
    \caption{Outage tolerance compliance }
    \label{fig:outage-compliance}
\end{figure}

{{Since $\rho$ is selected according to results obtained during the test week, it is important to analyze if the outage constraints have been enforced during the evaluation week, as depicted in Figure \ref{fig:outage-compliance}. We observe that those have been close to the imposed $\gamma$ thresholds, but failed to meet them in several cases. Although no severe errors arise, the violation of the outage constraints highlights the need to implement re-training schemes. Network utilization patterns are not constant, and using only two weeks for training does not capture all the existing scenarios. Moreover, as longer periods elapse (months, years), higher drift will appear in the model with respect to the input data distribution, particularly in cellular networks, where the continuous increase of 5G traffic will directly influence the existing switch-off opportunities. 

{To evaluate the cost of using a generalized model, we compare the obtained results against our previous approach \cite{mswim-paper}, where an XGBoost Classifier was used as the on/off decision algorithm. We recompute the average switch-off times matrix (Figure \ref{fig:achievable-savings}) for the previous strategy, and then get the difference in the off times obtained. Recall that, the XGBoost-based approach requires training one model per $(T, \gamma)$ tuple, which leads to a total use of eighteen fine-tuned models. When using the LSTM-based approach, we observe an average reduction of only 2.03\% of identified switch-off opportunitiess. Specifically, the average reduction at each $T$ across the three $\gamma$ constraints is equal to 1.4\%, 1.67\%, 3\%, 3.4\%, 0.6\%, and 1.2\% from 0 to 25 Mbps, respectively. On the other hand, the average reduction at each $\gamma$ across all $T$ levels is equal to 0.4\%, 2.4\%, and 2.9\% for $\gamma = 10\%, 5\%, \text{ and }3\%$, respectively. 

In general, we observe that the cost of using a generalized model is not severe, and instead reduces the overall complexity and maintenance of the proposed solution. Therefore, the LSTM-based solution effectively overcomes the scalability limitations observed in the XGBoost approach, positioning it as a practical alternative for operators seeking to deploy adaptive QoS-aware energy-saving strategies. }

\subsection{Energy footprint and economic benefit: an operator's perspective}

{The above results demonstrate that potential energy savings exist even under stringent QoS constraints. To analyze the benefits our strategy would have from an operator's perspective, we mapped the obtained results into CO\textsubscript{2} emissions reduction, and the corresponding economic benefits derived from operational expenses reduction by switching off cells. Recall that up to this point, we have worked with a subset of 70 cells and analyzed the energy saving strategy over one week of data. The results presented in this section are an extrapolation to large-scale deployments, which involve thousands of cells, and a full year of operation (i.e., 52 weeks). We estimate that the total amount of deployed 5G cells belonging to the same carrier as the ones we analyzed in the whole country for one operator is around five thousand.   }

{The energy consumption of a given cell can be decomposed into two main parts: the baseline consumption, and the consumption due to traffic demand. This means that offloading traffic from the 5G cells to the 4G cells will increase the consumption of the latter ones. Unfortunately, the MNO's dataset only provided energy KPIs related to 5G nodes, but not to 4G ones. Hence, as we did in \cite{previous-paper}, we estimate the energy savings as the baseline consumption of the 5G cells each time those are switched off. Concretely, this represents 133 Wh per 15-minute switch-off period.  }

{On the one hand, mapping into equivalent kg of CO\textsubscript{2} emissions is done by using the conversion factor of 0.26 kg/kWh stated by the national regulatory entity of the analyzed deployment country. On the other hand, since the electricity price varies during the day, mapping to economic benefits requires examining the switch-off hours distribution, which is shown in Figure \ref{fig:off-hours}. We observe that most of the switch-off decisions are taken during nighttime (approximately 70\% of the switch-off decisions are taken between 9pm and 7am). Of course, if we would separately look at the switch-off hours distribution of each ($T$, $\gamma$) tuple, we would observe that the higher the constraints, the fewer switch-off decisions are made during day time. 

\begin{figure}[t]
    \centering    \includegraphics[width=0.42\textwidth]{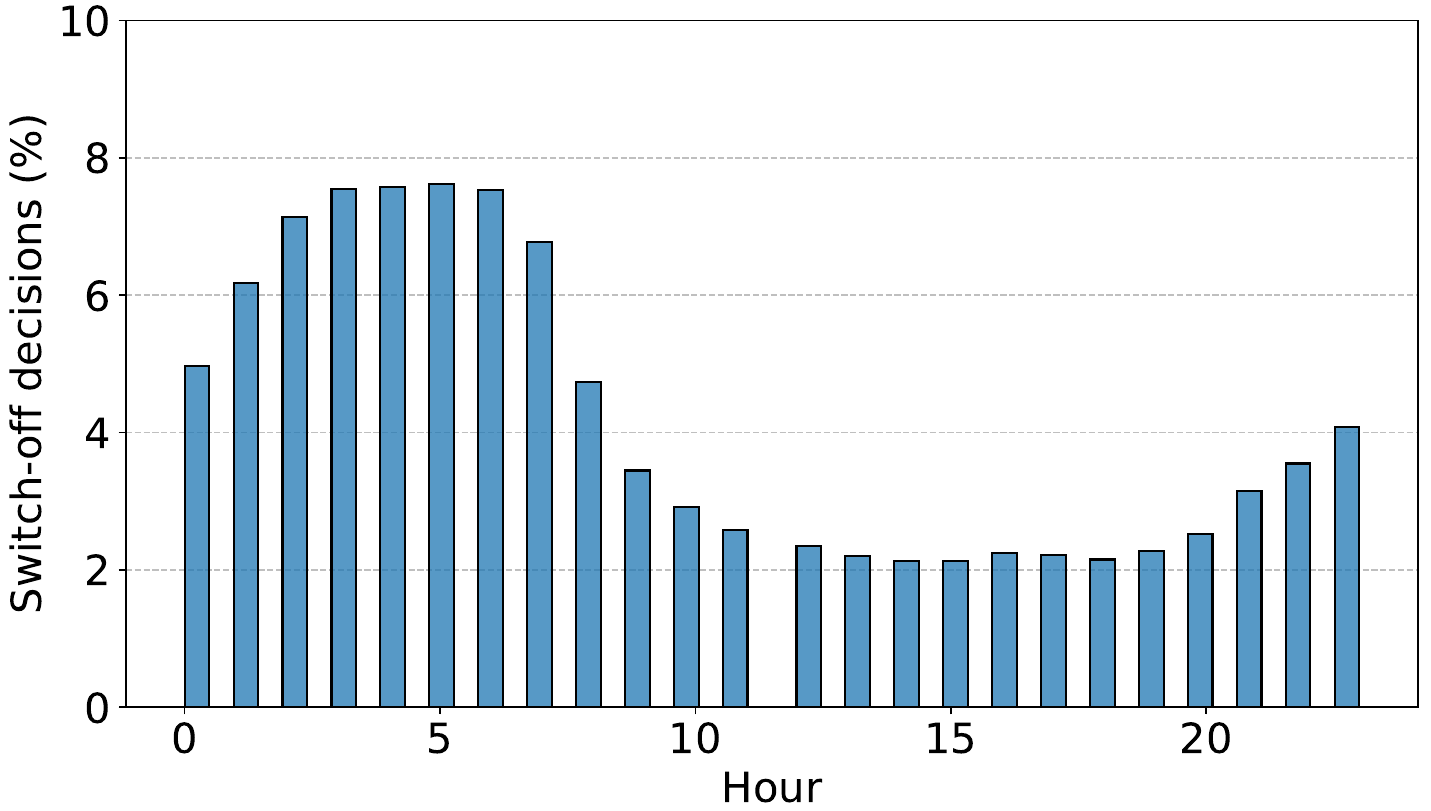}
    \caption{Switch-off hours distribution.}
    \label{fig:off-hours}
\end{figure}

{To finally map the energy savings into economic benefit, we look again at regulatory agencies to get a conversion factor between MWh and euros. Concretely, in this country we get an average value of 90 - 100 \euro/MWh during nighttime, increasing during day time. To compute our estimations we select an average price of 95 \euro/MWh. Table \ref{tab:mappings} provides the estimation on the energy savings in GWh, tonnes of CO\textsubscript{2} emissions reduction, and economic benefit estimation in millions of euros for six representative tuples of soft, intermediate, and strong QoS constraints, obtained across five thousand cells and a full year of operation.

\begin{table}[t]
\caption{Estimated operator benefits per year. }
\centering
\begin{tabular}{|c|c|c|c|}
\hline
\textbf{($T$, $\gamma$)} & \makecell{Energy \\ Savings \\(GWh)} & \makecell{CO\textsubscript{2} \\ emissions \\(tonnes)} & \makecell{Economic \\benefit \\ (M\euro)} \\
\hline
(5 Mbps, 10\%) & 16.5 & 4290 & 1.67 \\
\hline
(15 Mbps, 10\%) & 10.9 & 2834 & 1.04 \\
\hline
(25 Mbps, 10\%) & 5.3 & 1378 &  0.5 \\
\hline
(5 Mbps, 3\%) & 11.1 & 2886 & 1.06 \\
\hline
(15 Mbps, 3\%) & 7.9 & 2054 &  0.75 \\
\hline
(25 Mbps, 3\%) & 4.1 & 1066 &  0.39 \\
\hline
\end{tabular}
\label{tab:mappings}
\end{table}

{Overall, lower throughput levels and relaxed outage policies lead to greater energy savings, CO\textsubscript{2} reductions, and economic gains. For instance, with a level of 5 Mbps and $\gamma=10\%$, the proposed strategy achieves 16.5 GWh of annual energy savings, corresponding to 4290 tonnes of avoided CO\textsubscript{2} emissions, and an economic benefit of 1.67 M\euro. Conversely, increasing the throughput requirement to 25 Mbps while keeping $\gamma=10\%$ reduces the benefits by nearly a factor of three. Although outage tolerance also plays a decisive role (for $T=5$~Mbps, decreasing $\gamma$ from 10\% to 3\% reduces energy savings from 16.5 to 11.1 GWh and the economic impact from 1.67 to 1.06 M\euro), increasing the throughput level results in a more significant reduction of the achievable benefits. 


%% file: conclusions.tex
\section{Conclusions}\label{sec:conclusions}


{

This paper presented a QoS-aware cell switch-off strategy that combines a temporal Long Short-Term Memory (LSTM) model with an inference-time decision threshold~\(\rho\) to enforce operator policies specified by target throughput~\(T\) and outage tolerance~\(\gamma\). By calibrating \(\rho\) on a testing week and fixing it during evaluation, the approach enables policy control without retraining, avoiding the model replication inherent to custom-loss or class-ratio tuning. On an unseen week and across multiple \(T\) levels, the method achieves \(63\text{--}96\%\) of the oracle’s energy-saving upper bound while meeting the specified outage constraints, showing that simple post-hoc thresholding can turn a single trained model into a practical and operator-tunable controller. 

Beyond technical validation, we translate switch-off time into deployment-level impact, reporting indicative CO\(_2\) and OPEX figures under stated assumptions. These estimates are intended to illustrate business relevance rather than to serve as definitive accounting; they reflect {5G baseline savings only} (4G backfill energy is not included due to missing KPIs) and should therefore be interpreted as upper bounds that will tighten with richer energy models.

The results suggest a straightforward path to O-RAN integration (e.g., rApp/SMO): periodic recalibration of \(\rho\) can maintain policy compliance as traffic patterns evolve, with retraining reserved for drift or feature updates.
Future work will also include:
\begin{itemize}
  \item Leverage O-RAN digital-twin implementations to enable broader  validation of the proposed mechanisms.
  \item Automate weekly calibration and monitoring of \(\rho\).
  \item Extend constraints beyond scalar \(T\) and \(\gamma\) to multi-metric QoS (e.g., latency, percentile throughput).
  \item Incorporate 4G backfill consumption and richer power models to refine CO\(_2\)/OPEX estimates.
\end{itemize}